\documentclass[useAMS,usenatbib]{mn2e}
\usepackage{graphicx,ifthen,dcolumn,url}

\def\aatab#1#2#3#4#5#6{\ifthenelse{\equal{*}{#1}}
{\begin{table*}\caption[]{\label{#2} #3}
  \begin{flushleft}
    \begin{tabular}{#4}
      \hline\noalign{\smallskip} #5
      \noalign{\smallskip} \hline \noalign{\smallskip} #6
      \noalign{\smallskip} \hline
    \end{tabular}
  \end{flushleft}
\end{table*}}
{\begin{table}\caption[]{\label{#1} #2}
  \begin{flushleft}
    \begin{tabular}{#3}
      \hline\noalign{\smallskip} #4
      \noalign{\smallskip} \hline \noalign{\smallskip} #5
      \noalign{\smallskip} \hline
    \end{tabular}
  \end{flushleft}
\end{table}}}

\newcommand{\be}[1]{\begin{equation}\label{#1}}
\newcommand{\ee}{\end{equation}}
\newcommand{\bea}[1]{\begin{eqnarray}\label{#1}}
\newcommand{\eea}{\end{eqnarray}}
\newcommand{\lwig}{{\leavevmode\kern0.3em\raise.3ex\hbox{$<$}
                    \kern-0.8em\lower.7ex \hbox{$\sim$}\kern0.3em}}
\newcommand{\dd} [2]{{{{\rm d}{#1}\over{\rm d}{#2}}}}

\renewcommand{\dd}{{\rm d}}

\newlength{\pwidth}
\newcommand{\ppmm}[2]{\lower-.7ex\hbox{\scriptsize$#1$}\settowidth{\pwidth}%
            {\scriptsize$#2$}\kern-\pwidth\lower.5ex\hbox{\scriptsize$#2$}}
\def\note #1]{{\bf #1]}}

\def\Msun{\, {\rm M}_\odot}

\def\mdis{m_{\rm dis}}
\def\qdis{q_{\rm dis}}
\def\rdis{r_{\rm dis}}

\def\qcmax{q_{\rm c,max}}
\def\qeps{q_\epsilon}
\def\mshell{m_{\rm shell}}
\def\rshell{r_{\rm shell}}
\def\CA{{\cal A}}
\def\CB{{\cal B}}
\def\clight{\tilde c}
\def\mua{\mu_{\rm a}}
\def\mub{\mu_{\rm b}}
\def\pbcz{p_{\rm bcz}}
\def\Tbcz{T_{\rm bcz}}
\def\rbcz{r_{\rm bcz}}
\def\rbf{\rm}

\begin{document}
    \title[On the red-giant luminosity bump]
          {On the red-giant luminosity bump}
    \author[J. Christensen-Dalsgaard]{J{\o}rgen Christensen-Dalsgaard$^{1,2}$\thanks{E-mail: jcd@phys.au.dk}\\
    $^{1}$Stellar Astrophysics Centre, Department of Physics and Astronomy,
          Aarhus University, DK--8000 Aarhus C, Denmark\\
    $^{2}$Kavli Institute for Theoretical Physics, Kohn Hall,
University of California Santa Barbara, CA 93106, USA
}

    \date{Received 4 June 2015 / Accepted \today}
    \pagerange{\pageref{firstpage}--\pageref{lastpage}} \pubyear{2013}

    \maketitle

    \label{firstpage}

    \begin{abstract}
The increase in luminosity as a star evolves on the red-giant branch is
interrupted briefly when the hydrogen-burning shell reaches the vicinity
of the composition discontinuity left behind from the first convective 
dredge-up.
The non-monotonic variation of luminosity causes an accumulation of stars,
known as the `bump', in the distribution of stars in the
colour-magnitude diagrams of stellar clusters,
which has substantial diagnostic potential.
Here I present numerical results on this behaviour and discuss
the physical reason for the luminosity variation, with the goal of 
strengthening the understanding of origin of the phenomenon and hence of its
diagnostic potential.
    \end{abstract}


\section{Introduction}
\label{sect:intro}



\begin{figure}
\centering
\includegraphics[width=\hsize]{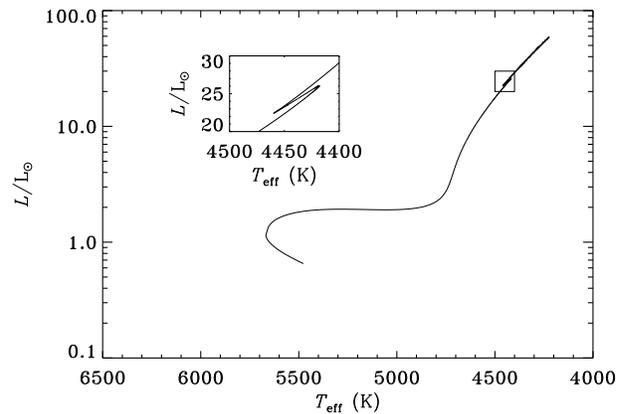}
\caption{Evolution track of $1 \Msun$ model.
The inset shows the vicinity of the luminosity bump, marked by the box.
\label{fig:HR}
}
\end{figure}

During the evolution up the red-giant branch models of low-mass stars show
a brief phase of decreasing luminosity (cf. Fig.\,\ref{fig:HR}).
This was early \citep{Thomas1967, Iben1968} associated with the 
approach of the hydrogen-burning shell to the discontinuity in the hydrogen
abundance left behind by the first dredge-up, where the convective envelope
reaches into parts of the star earlier affected by hydrogen fusion during
main-sequence evolution.
As a result of this luminosity variation the star obviously spends
a little longer in the relevant region of the Hertzsprung-Russell diagram.
In stellar clusters this corresponds to an accumulation of stars in that 
region, leading to a `bump' in the luminosity distribution 
\citep{Iben1968, King1985};
consequently the phenomenon is known as the `red-giant bump'.%
\footnote{Not to be confused with the `red-giant clump' consisting of stars 
in the core helium burning phase.}
The location in luminosity of the bump and the excess number of
stars are clearly closely related to a specific phase in
stellar evolution, making it an interesting diagnostics for 
the properties of the clusters and/or the stellar modelling
\citep[e.g.,][]{Fusi1990, Cassis1997, Bono2001, Riello2003, Nataf2014}.
A particularly interesting aspect is the effect of overshoot from the
convective envelope, which obviously increases the depth of the composition
discontinuity and hence decreases the luminosity at which the bump is found
{\rbf \citep[e.g.,][]{Alongi1991, Angelo2015}.}
For a detailed overview of the diagnostic potential, and further references,
see \citet{Salari2002}.

The study of red giants has recently received an enormous boost through
the asteroseismic analyses made possible by the photometric observations from 
the CoRoT \citep{Baglin2009} and {\it Kepler} \citep{Boruck2010} 
space missions, motivating further investigations of the properties
of the bump.
An example is the signature brought about in the observed frequencies 
by buoyancy glitches,
which appears to be related to the evolutionary location of the star relative
to the bump \citep{Cunha2015}.


The detailed reason for the variation in luminosity has been subject of
some discussion. 
It has been associated with the direct effect of the increase
in the hydrogen abundance in the hydrogen-burning shell
as it crosses the composition discontinuity,
possibly involving departure from thermal equilibrium
\citep[e.g.,][]{King1985, Salari2002, Riello2003, Gai2015}.
However, in a careful presentation of the evolution in the vicinity of the bump
\citet{Sweiga1990} noted that since the decrease in luminosity happens
{\it before} the hydrogen-burning shell reaches the composition 
discontinuity the effect cannot be due to the increase in the fuel
available to the nuclear reactions;
instead they pointed to the increase in the opacity just outside the shell
due to the increased hydrogen abundance in the envelope.
A perhaps more likely interpretation was obtained by \citet{Refsda1970}
\citep[see also][]{Kippen1990} who carried out a homology analysis of the 
properties of the region just outside the burning shell.
They noted a strong increase in the luminosity with the mean molecular 
weight $\mu$ and pointed out that the increase in the hydrogen abundance
outside the discontinuity caused a decrease in $\mu$ and hence in luminosity.

Here I illustrate the properties of the evolution in the vicinity of the bump
in a little more detail, based on a series of evolution sequences of
varying mass.
In addition, I carry out a simplified analysis of the structure of the model
just above the hydrogen-burning shell which essentially confirms the 
importance of the variation in mean molecular weight as the driving force
behind the bump.

\begin{figure}
\centering
\includegraphics[width=\hsize]{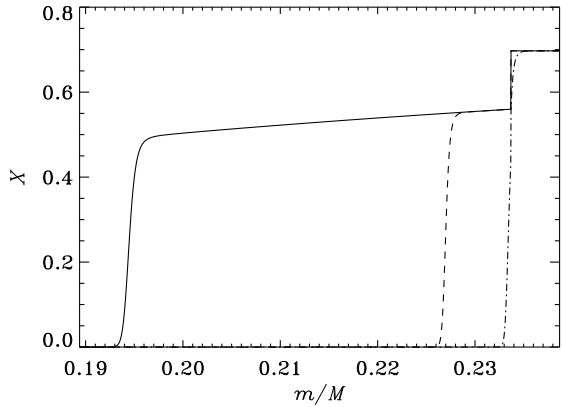}
\caption{Hydrogen abundance as a function of fractional mass, 
in a $1 \Msun$ evolution sequence, at the times of
maximum extent of the convective envelope (solid),
and at luminosity maximum (dashed) and minimum (dot-dashed) in the bump.
\label{fig:xprof}
}
\end{figure}


\section{Properties of the bump}


The models considered here follow the computations of \citet{Jiang2014}.
Briefly, they use the ASTEC code \citep[][]{Christ2008}, neglecting 
diffusion and settling and with no overshoot from the convective envelope
and possible convective core.
The models have an initial hydrogen abundance $X_0 = 0.72$ and a heavy-element
abundance $Z = 0.02$; convection is treated with the \citet{Bohm1958}
mixing-length formulation with a mixing-length parameter of 1.8.

In the early phases of red-giant evolution the mass in the
convective envelope increases to a maximum and subsequently decreases.
The convective envelope is assumed to be fully mixed, leading to 
the {\it first dredge-up} of nuclear-processed material,
down to the maximum extent of the convective envelope
where the mass fraction at its base is $\qdis = \mdis/M$,
$\mdis$ being the mass inside the discontinuity and $M$ the total mass
of the star.
Owing to the neglect of diffusion and settling the composition at the
base of the mixed region is discontinuous.
Figure\,\ref{fig:xprof} shows the hydrogen abundance as a function of
fractional mass in the $1 \Msun$ sequence, at the times of the maximum
extent in mass of the convective envelope, the maximum luminosity in the bump,
and the time where the location, at a fractional mass $\qeps$,
of the hydrogen-burning shell reaches the composition discontinuity;
here $\qeps$ is defined by the maximum in the energy-generation rate.
The latter time corresponds closely to the minimum in luminosity in the bump.

\begin{figure}
\centering
\includegraphics[width=\hsize]{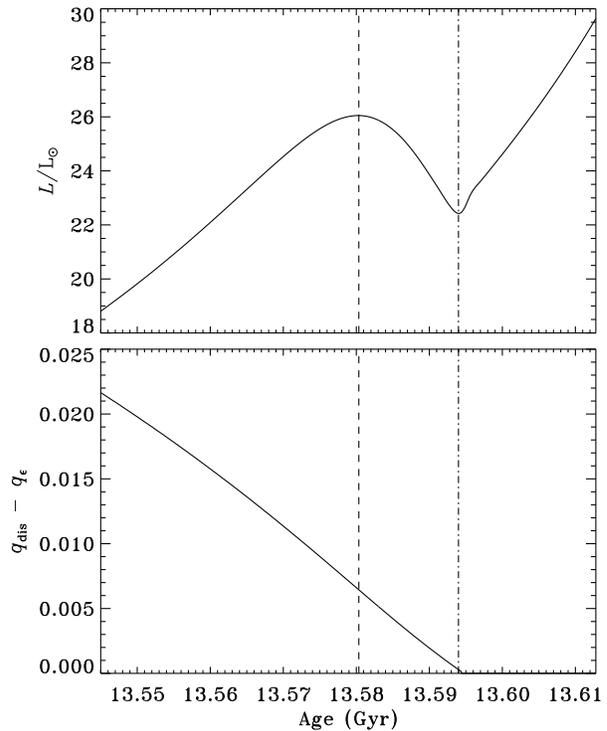}
\caption{The top panel shows the variation of surface luminosity 
in the vicinity of the bump, in solar units, as a function of age
in a $1 \Msun$ evolution sequence.
The lower panel shows the separation in mass fraction between the location
of the composition discontinuity and the hydrogen-burning shell.
In both panels the dashed and dot-dashed lines, as in
Fig.\,\ref{fig:xprof}, mark the luminosity maximum and minimum respectively.
\label{fig:lumi}
}
\end{figure}

The evolution with time in the luminosity in the vicinity of the bump 
is illustrated in Fig.\,\ref{fig:lumi}.
The lower panel shows the separation between $\qeps$ and $\qdis$, 
again making clear that the luminosity starts to increase as soon as the 
hydrogen-burning shell crosses the discontinuity.
{\rbf Here and in the following I ignore the fact that the age
of some of the models considered,
with the assumed parameters, exceeds the inferred age of the Universe.
This has no significance for the properties of the models.}

\begin{figure}
\centering
\includegraphics[width=\hsize]{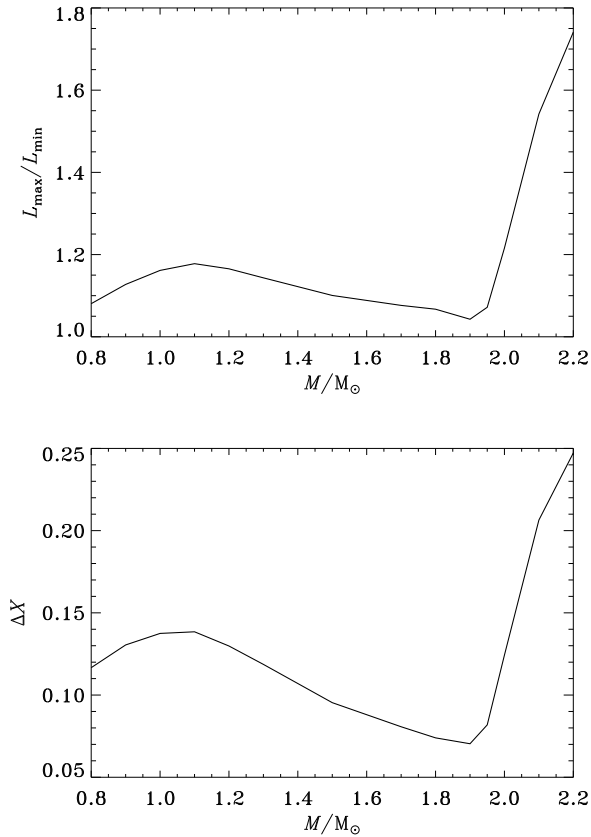}
\caption{The top panel shows the range of the bump, in terms of the 
ratio between the maximum and minimum luminosity,
as a function of stellar mass.
The bottom panel shows the jump $\Delta X$ in hydrogen abundance across
the dredge-up discontinuity (cf. Fig.\,\ref{fig:xprof}).
\label{fig:lumirat}
}
\end{figure}



The properties of the bump depend substantially on stellar mass
\citep[see also][]{Gai2015}.
This is illustrated in Fig.\,\ref{fig:lumirat}, showing the ratio between
the maximum and minimum luminosity in the bump.
As shown by the lower panel, this is very strongly correlated with the
step $\Delta X$ in hydrogen abundance at the discontinuity.
Note that the red-giant bump only appears for stars of mass below around
$2.2 \Msun$. 
For more massive stars helium ignition occurs before the hydrogen-burning shell
reaches the composition discontinuity.
In such stars the bump can be identified during the helium-burning phase
\citep[e.g.,][]{Cunha2015}.

\begin{figure}
\centering
\includegraphics[width=\hsize]{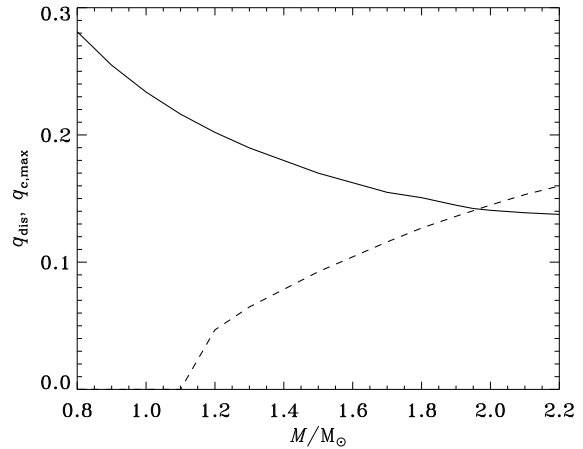}
\caption{The solid line shows the fractional mass $\qdis$ at the composition
discontinuity and the dashed line shows the maximum extent $\qcmax$ of the
convective core, as functions of stellar mass.
\label{fig:qmin}
}
\end{figure}



\begin{figure}
\centering
\includegraphics[width=\hsize]{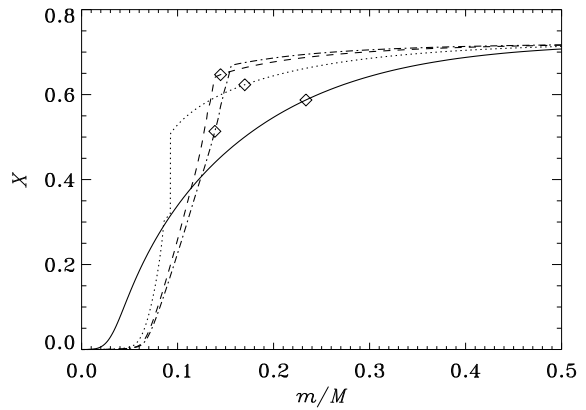}
\caption{Hydrogen profile at the end of central hydrogen burning,
as a function of fractional mass, for stars of
mass $1 \Msun$ (solid), $1.5 \Msun$ (dotted), $1.9 \Msun$ (dashed)
and $2.1 \Msun$ (dot-dashed).
The diamonds mark the subsequent location of the maximum extent $\qdis$
of the dredge-up.
\label{fig:hydroprof}
}
\end{figure}

The variation of $\Delta X$ with mass arises from a somewhat complex
interaction between the extent of the convectively mixed regions and
main-sequence nuclear burning.
Figure\,\ref{fig:qmin} shows the variation of the mass fraction $\qdis$
at the discontinuity
and the maximum extent $\qcmax$ of the mass fraction in the convective core.
With increasing stellar mass 
the maximum penetration in mass of the convective envelope increases,
and hence $\qdis$ decreases, monotonically.
At the lowest masses this causes a slight increase in $\Delta X$ with mass
(cf. Fig.\,\ref{fig:lumirat}),
but above around $1 \Msun$ the decrease in $\qdis$ is more than 
compensated by the increasing central concentration of hydrogen burning
as the CNO cycle becomes more important.
This is illustrated in Fig.\,\ref{fig:hydroprof} which shows the 
hydrogen-abundance profile at the end of central hydrogen burning,
for various stellar masses.
For masses higher than around $1.2 \Msun$ the models have a convective core
whose maximal extent $\qcmax$ increases rapidly with mass.
When $\qcmax$ crosses $\qdis$, around $1.9 \Msun$, the steep
hydrogen profile left behind by the retreating convective core leads to
a rapid increase in $\Delta X$ and hence in the amplitude of the bump.

I note in passing that, even though the phase associated with the bump
is relatively brief, it does not lead to any substantial departure
from thermal equilibrium;
{\rbf such departure was implied, e.g., by \citet{Riello2003}.}
In the $1 \Msun$ sequence, for example, the contribution to the total
luminosity from gravitational effects peaks at a value slightly less
than 1 per cent around the time of minimum luminosity during the bump.
In general, the dominant gravitational effect appears to be associated 
with the region of shell burning, even far away from the bump. 

\section{Analysis of the bump}


According to the shell homology analysis by \citet{Refsda1970} the temperature
$T$ at homologous points scales as
\begin{equation}
T \sim \mu {\mshell \over \rshell} \; ,
\label{eq:simhom}
\end{equation}
where $\mu$ is the mean molecular weight, and $\mshell$ and $\rshell$ are 
the mass and radius at the hydrogen-burning shell source.
The increasing mass of the helium core causes an increase in temperature and
hence in luminosity, owing to the strong temperature sensitivity of the
energy generation rate and efficiency of the energy transport.
This assumes a constant mean molecular weight in the region considered,
although \citet{Refsda1970} noted the potential effect of the decrease in
the average $\mu$, leading to a decrease in luminosity,
as the shell source approaches the composition discontinuity
resulting from the dredge-up.
Here I analyse in more detail the effect of the discontinuity on the structure
of this region.

As did \citet{Refsda1970} I assume the ideal gas law, so that pressure $p$,
density $\rho$ and $T$ are related by
\begin{equation}
p = {k_{\rm B} \rho T \over \mu m_{\rm u}} \; ,
\label{eq:eos}
\end{equation}
where $k_{\rm B}$ is Boltzmann's constant and $m_{\rm u}$ is the atomic
mass unit.
From the equations of hydrostatic equilibrium and radiative energy transport
we obtain
\begin{equation}
{\dd T^4 \over \dd p} = {3 \kappa L \over 4 \pi a \clight G m} \; ,
\label{eq:radgrad}
\end{equation}
where $\kappa$ is the opacity, $a$ the radiation density constant,
$\clight$ the speed of light, $G$ the gravitational constant and $m$
the mass inside the given point.
I consider the region between the shell source and the base of the
convective envelope,
neglecting the mass in the region outside the shell, so that $m \simeq \mshell$.
Outside the shell I also take $L$ to be constant.

The treatment of the opacity requires a little care.
As implicitly noted by \citet{Sweiga1990} the opacity generally scales
with the number density of electrons, i.e., as $1 + X$, which in itself
would lead to an increase in the opacity moving across the discontinuity.
However, the bound-free and free-free contributions to the opacity in addition
are proportional to density.
Since $p$ and $T$ are continuous, $\rho$ scales as $\mu$ and hence 
{\it decreases} across the discontinuity, and so therefore do these 
opacity contributions.
In the models considered here the net result is a small decrease in opacity
across the discontinuity which, for simplicity, I ignore in the following.

As did \citet{Refsda1970} I approximate the opacity as
\begin{equation}
\kappa \simeq \kappa_0 p^a T^b \;,
\end{equation}
where, according to the argument just given, $\kappa_0$ is assumed to be
constant across the discontinuity.
From equation (\ref{eq:radgrad}) it then follows that
\begin{equation}
{\dd T^4 \over \dd p} = \CA p^a T^b \; ,
\end{equation}
where 
\begin{equation}
\CA = 
{3 \kappa_0 L \over 4 \pi a \clight G \mshell} \; .
\end{equation}
The solution of this equation should be matched to the temperature $\Tbcz$
and the pressure $\pbcz$ at the base of the convective envelope, where the
temperature gradient changes to being nearly adiabatic.
For simplicity I neglect $\Tbcz$ and $\pbcz$ compared with the values around
the discontinuity, to obtain
$T = \CB p^\gamma$, where $\gamma = (a+1)/(4 - b)$ and
\begin{equation}
\CB = \left( {\CA \over 4 \gamma} \right)^{1/(4 - b)} \; .
\end{equation}

Introducing $z = 1/r$, where $r$ is the distance to the centre,
the equation of hydrostatic support is approximated as
\begin{equation}
{\dd p \over \dd z} = G \mshell \rho \; ,
\end{equation}
or, using equation (\ref{eq:eos}) and $T \propto p^\gamma$,
\begin{equation}
{\dd T \over \dd z} = \gamma {T \over p} {\dd p \over \dd z} 
= \gamma G \mshell {\mu m_{\rm u} \over k_{\rm B}}\; .
\label{eq:hydeqt}
\end{equation}
I take $\mu = \mua$ above the discontinuity and $\mu = \mub$ below the 
discontinuity;
in the latter case I neglect the small variation, due to the gradient
in the hydrogen abundance, in the relatively thin region considered, so that
both $\mua$ and $\mub$ are taken to be constant.
Furthermore I assume that there is a point $r_0$ with
$\rshell \ll r_0 \ll \rbcz$, 
where 
$\rbcz$ is the distance of the base of the convective envelope from the centre,
such that equation (\ref{eq:hydeqt}) is valid
for $z \ge 1/r_0$.
Integrating equation (\ref{eq:hydeqt}) downwards from $z = 1/r_0$ yields,
for $r < \rdis$,
\begin{eqnarray}
\label{eq:tapprox}
T &=& {\gamma G m_{\rm u} \mshell \over k_{\rm B}}
\left( \int_{1/r_0}^{1/\rdis} \mua \dd z +
\int_{1/\rdis}^{1/r} \mub \dd z \right) \nonumber \\
&\simeq& {\gamma G m_{\rm u} \over k_{\rm B}} \mub {\mshell \over r}
\left [1 - {r \over \rdis} \left(1 - {\mua \over \mub} \right) \right] \; ,
\end{eqnarray}
where I neglected $T(r_0)$ and $1/r_0$ in the second line;
here $\rdis$ is the distance of the discontinuity from the centre.
Thus, with $r \simeq \rshell$ we recover the homology scaling in
equation (\ref{eq:simhom}) but with the {\rbf correction factor 
\begin{equation}
\phi = 1 - (\rshell / \rdis) [1 - (\mua / \mub) ]
\label{eq:corfact}
\end{equation}
which becomes significant} as $\rshell$ approaches $\rdis$,
reducing $T$ and hence the luminosity.

\begin{figure}
\centering
\includegraphics[width=\hsize]{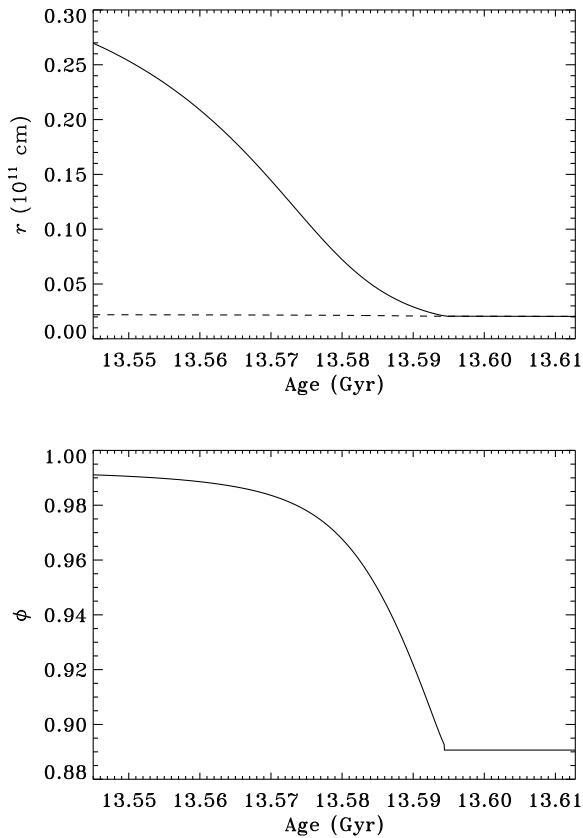}
\caption{The top panel shows the distance $\rdis$ of the discontinuity 
from the centre (solid) and the radius $\rshell$ 
of the hydrogen-burning shell (dashed),
as a function of age, in a $1 \Msun$ evolution sequence.
The bottom panel shows the corresponding {\rbf correction factor
$\phi$ (cf. equation \ref{eq:corfact}).}
After the minimum in luminosity $\rdis$ is replaced by $\rshell$,
and $\phi$ is set to $\mua/\mub$.
\label{fig:rdisc}
}
\end{figure}


To illustrate the properties of the correction induced by the discontinuity,
the upper panel in Fig.\,\ref{fig:rdisc} shows $\rdis$ and 
$\rshell$, 
while the lower panel shows the corresponding correction factor, 
in the $1 \Msun$ model.
It is obvious that the general shape of the correction factor matches the
reduction of $L$ shown in Fig.\,\ref{fig:lumi},
keeping in mind that the energy generation and transport depend on 
temperature to a high power.
A quantitative comparison with the model properties shows substantial 
departures from the simple expressions;
this is true both of the homology scaling in equation (\ref{eq:simhom})
well below the bump and
the solution in equation (\ref{eq:tapprox}) in the vicinity of the bump.%
\footnote{On the other hand, the scaling in equation (\ref{eq:simhom})
becomes increasingly well
satisfied for more luminous models on the red-giant branch.}
This is likely a result of the
rather drastic approximations made in the derivation in the latter equation,
particularly the neglect of quantities at the base of the convective envelope.
Even so I find it plausible, as implied by \citet{Refsda1970}, 
that the cause of the luminosity 
variation in the bump is predominantly the result of the effect of
the composition discontinuity on the hydrostatic
equilibrium, the effective weight of the overlying material being reduced
by the decrease in the mean molecular weight above the discontinuity.

\section{Discussion and conclusion}


The results presented in the present note relate the luminosity excursion in 
the red-giant bump directly to the jump in mean molecular weight at the
dredge-up composition discontinuity, as a consequence of its effect
on the hydrostatic structure of the region immediately above the
hydrogen-burning shell. 
The amplitude of the luminosity excursion, and hence the over-density of stars
in this region of the Hertzsprung-Russell diagram, is closely correlated
with the magnitude $\Delta X$ of the jump in the hydrogen abundance,
which in turn depends on the main-sequence evolution of the star;
this includes, in a rather narrow range of stellar masses, the maximum extent
of the convective core.
From this insight, possibly enhanced through more careful analysis, one may
hope to get a better feel for the diagnostic potential of the observed bumps.


Here I have considered the simple case of models neglecting diffusion and
settling and with no overshoot from the convective envelope.
Further investigations of how such effects might affect the bump would
clearly be interesting.
This includes studies of the effect on the observed oscillation frequencies
through the glitch in the buoyancy frequency \citep[e.g.,][]{Cunha2015}.
Convective overshoot and turbulent diffusive mixing near the base of the
convective envelope will certainly influence the location and
shape of the composition structure resulting from the dredge-up, which may well
have effects that can be inferred from asteroseismic analyses and which may
also be detectable in the luminosity-distribution bump for stellar clusters.

\section*{Acknowledgements}
I am grateful to V. Silva Aguirre, S. Hekker and G. Houdek,
as well as the entirely anonymous referee,
for useful discussions or comments on an earlier version of the manuscript.
I thank the Kavli Institute for Theoretical Physics for hospitality during
part of this work and the organizers of the 
KITP Program: Galactic Archaeology and Precision Stellar Astrophysics
for providing the opportunity for the visit.
Funding for the Stellar Astrophysics Centre is provided by 
the Danish National Research Foundation (Grant DNRF106).
The research is supported by the ASTERISK project
(ASTERoseismic Investigations with SONG and Kepler)
funded by the European Research Council (Grant agreement no.: 267864).
This research was supported in part by the National Science Foundation
under Grant No. NSF PHY11-25915, and
made use of NASA's Astrophysics Data System.

\label{lastpage}


\begin{thebibliography}{99}


\bibitem[Alongi {\rm et~al.\/}(1991)]{Alongi1991}
Alongi M., Bertelli G., Bressan A., Chiosi C., 1991,
{A\&A}, {\rm 244}, 95  

{\rbf
\bibitem[Angelou {\rm et~al.\/}(2015)]{Angelo2015}
Angelou G. C., D'Orazi V., Constantino T. N., Church R. P., 
Stancliff R. J., Lattanzio J. C., 2015,
{MNRAS}, {\rm 450}, 2423 
}

\bibitem[Baglin {\rm et~al.\/}(2009)]{Baglin2009}
Baglin A., Auvergne M., Barge P., Deleuil M., Michel E. and
the CoRoT Exoplanet Science Team, 2009,
in Pont F., Sasselov D., Holman M., eds,
{\rm Proc. IAU Symp. 253, Transiting Planets},
IAU and Cambridge University Press, p. 71  

\bibitem[B{\"o}hm-Vitense(1958)]{Bohm1958}
B{\"o}hm-Vitense E., 1958,
{\rm Z. Astrophys.}, {\rm 46}, 108  

\bibitem[Bono {\rm et~al.\/}(2001)]{Bono2001}
Bono G., Cassisi S., Zoccali M., Piotto G., 2001,
{ApJ}, {\rm 546}, L109  

\bibitem[Borucki {\rm et~al.\/}(2010)]{Boruck2010}
Borucki W.~J., Koch D., Basri G., et al., 2010,
{\rm Science}, {\rm 327}, 977  

\bibitem[Cassisi \& Salaris(1997)]{Cassis1997}
Cassisi S., Salaris M., 1997,
{MNRAS}, {\rm 285}, 593  

\bibitem[Christensen-Dalsgaard(2008)]{Christ2008}
Christensen-Dalsgaard J., 2008,
{Ap\&SS}, {\rm 316}, 13  

\bibitem[Cunha {\rm et~al.\/}(2015)]{Cunha2015}
Cunha M. S., Stello D., Avelino P. P., Christensen-Dalsgaard J., 
Townsend R. H. D., 2015, 
{\rbf {ApJ}, {\rm 805}, 127}

\bibitem[Fusi Pecci {\rm et~al.\/}(1990)]{Fusi1990}
Fusi Pecci F., Ferraro F. R., Crocker D. A., Rood R. T., 
Buonanno R., 1990,
{A\&A}, {\rm 238}, 95  

\bibitem[Gai \& Tang(2015)]{Gai2015}
Gai N., Tang Y., 2015,
{ApJ}, {\rm 804}, 6 

\bibitem[Iben(1968)]{Iben1968}
Iben I., 1968,
{\rm Nature}, {\rm 220}, 143  

\bibitem[Jiang \& Christensen-Dalsgaard(2014)]{Jiang2014}
Jiang C., Christensen-Dalsgaard J., 2014,
{MNRAS}, {\rm 444}, 3622  

\bibitem[King {\rm et~al.\/}(1985)]{King1985}
King C. R., Da Costa G. S., Demarque P., 1985,
{ApJ}, {\rm 299}, 674  

\bibitem[Kippenhahn \& Weigert(1990)]{Kippen1990}
Kippenhahn R., Weigert A., 1990,
{\rm Stellar structure and evolution},
Springer-Verlag, Berlin

\bibitem[Nataf(2014)]{Nataf2014}
Nataf D. M., 2014,
{MNRAS}, {\rm 445}, 3839  

\bibitem[Refsdal \& Weigert(1970)]{Refsda1970}
Refsdal S., Weigert A., 1970,
{A\&A}, {\rm 6}, 426  

\bibitem[Riello {\rm et~al.\/}(2003)]{Riello2003}
Riello M., Cassisi S., Piotto G., Recio-Blanco A., De Angeli F.,
Salaris M., Pietrinferni A., Bono G., Zoccali M., 2003,
{A\&A}, {\rm 410}, 553  

\bibitem[Salaris {\rm et~al.\/}(2002)]{Salari2002}
Salaris M., Cassisi S., Weiss A., 2002,
{PASP}, {\rm 114}, 375  

\bibitem[Sweigart {\rm et~al.\/}(1990)]{Sweiga1990}
Sweigart A. V., Greggio L., Renzini A., 1990,
{ApJ}, {\rm 364}, 527  

\bibitem[Thomas(1967)]{Thomas1967}
Thomas H.-C., 1967,
{\rm Z. Astrophys.}, {\rm 67}, 420  


\end{thebibliography}
\end{document}